\input{aipcheck}

\documentclass[
    ,final            
  ]
  {aipproc}

\layoutstyle{6x9}

\newcommand{\be}[1]{\begin{equation}\label{#1}}
\newcommand{\beq}{\begin{equation}}
\newcommand{\eeq}{\end{equation}}
\newcommand{\beqn}[1]{\begin{eqnarray}\label{#1}}
\newcommand{\eeqn}{\end{eqnarray}}

\renewcommand{\to}{\rightarrow}

\def\ov{\overline}
\def\ee{\end{equation}}

\def\lsim{\raise0.3ex\hbox{$\;<$\kern-0.75em\raise-1.1ex
\hbox{$\sim\;$}}}
\def\gsim{\raise0.3ex\hbox{$\;>$\kern-0.75em\raise-1.1ex
\hbox{$\sim\;$}}}
\def\cal{\mathcal}

\def\cO{{\cal O}}

\def\ga{\gamma}
\def\Ga{\Gamma}

\def\La{\Lambda}  

\def\Om{\Omega}

\def\tphi{\tilde{\phi}}

\def\tf{\tilde{f}}

\def\tq{\tilde{q}}
\def\tl{\tilde{l}}

\def\te{\tilde{e}}
\def\tu{\tilde{u}}
\def\td{\tilde{d}}
\def\tnu{\tilde{\nu}}

\def\lpr{l^\prime}
\def\phpr{\phi^\prime}

\def\barl{\bar{l}}
\def\barphi{\bar{\phi}}
\def\barlpr{\bar{l}^\prime}
\def\barphpr{\bar{\phi}^\prime}
\begin{document}

\title{ Marriage between the baryonic  and dark matters }


\author{ Zurab Berezhiani }{
  address={ Dipartimento di Fisica, Universit\`a di L'Aquila, 
I-67010 Coppito, AQ, Italy \\
and INFN, Laboratori Nazionali del Gran Sasso,
 I-67010 Assergi, AQ, Italy }
}



\begin{abstract}
The baryonic and dark matter fractions can be both generated simultaneously and with comparable amounts,  
if dark matter is constituted 
by the baryons of the mirror world, a parallel hidden sector with 
the same (or similar) microphysics as that of the observable world. 
\end{abstract}

\maketitle


\subsection{Baryonic and Dark Matter in the Universe: 
Why $\Om_{\rm B} \sim \Om_{\rm D}$?}

Cosmological observations indicate 
that the Universe is nearly flat, with the energy density very close 
to critical: $\Om_{\rm tot} = 1$. The non-relativistic matter 
in the Universe consists of two components, baryonic (B)  
and dark (D). 
The recent data fits imply  \cite{SDDS}: 
\be{Om}
\Om_{\rm B} h^2 = 0.0222 \pm 0.0007, ~~~~ 
\Om_{\rm D} h^2 = 0.105 \pm 0.004,  
\ee 
where $h=0.73\pm 0.02$ is the Hubble parameter. 
Hence, the matter gives only a smaller fraction of the total energy density: 
$\Om_{\rm M} = \Om_{\rm B} +\Om_{\rm D} = 0.24 \pm 0.02$,  
while the rest is attributed to dark energy (cosmological term): 
$\Om_\La = 0.76 \mp 0.02$ \cite{SDDS,PDG}.  

The closeness of $\Om_\La$ and $\Om_{\rm M} $  
($\Om_\La/\Om_{\rm M} = 3.2 \pm 0.3$),    
known as cosmic coincidence, may have an antrophic origin: 
the matter and vacuum energy densities 
scale differently with the expansion of the Universe:
$\rho_{\rm M} \sim a^{-3}$ and $\rho_\La \sim $ const., 
and thus they must coincide at some moment. 
So, it is our good luck to assist the epoch when $\rho_{\rm M} \sim\rho_\La$: 
in the earlier Universe one had $\rho_{\rm M} \gg \rho_\La$ and in the 
later Universe one will have $\rho_{\rm M} \ll \rho_\La$. 
Moreover, if $\rho_\La$ would be just few times bigger, 
no galaxies could be formed and thus there would be nobody to 
rise the question. 

The closeness between $\Om_{\rm D}$ and $\Om_{\rm B} $
($\Om_{\rm D}/\Om_{\rm B} = 4.7 \pm 0.3$) 
gives rise to more painful problem. 
Both $\rho_{\rm D}$ and $\rho_{\rm B}$ scale as $\sim a^{-3}$ 
with the Universe expansion, and their ratio should not  dependent on time. 
Why then these two fractions are comparable, 
if they have a drastically different nature and different origin?

The baryon mass density is $\rho_{\rm B} =m_{\rm B} n_{\rm B}$,     
where $m_{\rm B} \simeq 1$ GeV is the nucleon mass, and 
$n_{\rm B}$ is the baryon number density. 
Hence,  $\Om_{\rm B} h^2 = 2.6 \times 10^8 (n_{\rm B}/s)$, $s$ being the 
entropy density, and so eq. (\ref{Om}) translates in 
$Y_B= n_{\rm B}/s \approx 0.85\times 10^{-10}$, 
in a nice consistence with the Big Bang Nucleosynthesis (BBN) bounds 
$Y_B = (0.5-1) \times 10^{-10}$ \cite{PDG}.  
The origin of non-zero baryon asymmetry $Y_B$, which 
presumably was produced in a very early universe 
as a tiny difference $n_{\rm B} = n_b - n_{\bar b}$
between the baryon and anti-baryon abundances, is yet unclear.  
The popular mechanisms known as GUT Baryogenesis, Leptogenesis, 
Electroweak Bariogenesis, etc., all are conceptually based on 
out-of-equilibrium processes violating $B(B-L)$ and C/CP \cite{Sakh,KRS}, 
and they  generically predict $Y_B$   as a function of  
the relevant interaction strengths and CP-viollating phases.

Concerning dark matter, almost nothing is known besides the fact it must be 
constituted by some cold relics with mass $m_{\rm D}$ 
which exhibits enormous spread between different popular candidates    
as e.g. axion ($\sim 10^{-5}$ eV), 
sterile neutrino ($\sim 10$ keV), 
WIMP/LSP ($\sim 1$ TeV), or Wimpzilla  ($\sim 10^{14}$ GeV).  
Non of these candidates has any organic  
link with any of the popular baryogenesis schemes.
The respective abundances $n_{\rm D}$ could be produced 
thermally (e.g. freezing-out of WIMPs) or non-thermally  (e.g. axion condensation 
or gravitational preheating for Wimpzillas),  but in no case they are related 
to the CP-violating physics.    
In this view,  the conspiracy between $\rho_{\rm D} = m_{\rm D} n_{\rm D}$
and $\rho_{\rm B} =m_{\rm B} n_{\rm B}$ indeed looks as a big paradox.  
What could be at the origin of the mysterious cosmic {\it Fine Tuning} 
between various {\it ad hoc} independent parameters
related to different sectors of the particle physics and different epochs of cosmology?


\subsection{Ordinary and Mirror Worlds} 

The old hypothesis \cite{LY} that there may exists a mirror world,
a hidden parallel sector of particles and interactions which is
the exact duplicate of our observable world,  
has attracted a significant interest over the past years 
in view of interesting implications for the particle physics
and cosmology 
(see e.g. \cite{Alice,IJMPA-F} for reviews).  
Such a theory is based on the product $G\times G'$ of two 
identical gauge factors with identical particle contents, 
where ordinary (O) particles belonging to $G$ are singlets of $G'$, 
and mirror (M) particles belonging to $G'$ are singlets of $G$.
Mirror parity under the proper interchange of  $G\leftrightarrow G'$
and the respective matter fields \cite{FLV} renders the  
Lagrangians of two sectors identical. 
The two worlds can be viewed as parallel branes in a higher dimensional space,
with O-particles localized on one brane and the M-particles on another brane,
while gravity propagates in the bulk. 
Such a setup can be realized in the  string theory context.

Besides gravity, two sectors could communicate by other means. 
In particular, any neutral O-particle, elementary or composite,   
could have a mixing with its M-twin.  
E.g. kinetic mixing of ordinary and mirror photons  \cite{Holdom}, 
mass mixing between ordinary and mirror neutrinos 
~\cite{FV} and neutrons \cite{nn}, etc. 
Such mixings may be induced by effective interactions between 
O- and M-particles mediated  by messengers between two sectors,  
e.g. gauge singlets  or some extra gauge bosons interacting with 
both sectors
~\cite{PLB98}. 

If the mirror sector exists, then the Universe 
should contain along with the ordinary photons, electrons, nucleons 
etc., also their mirror partners with exactly the same masses and 
exactly the same microphysics.  
%
However, two sectors must have different cosmological evolutions: 
in particular, they never had to be in equilibrium with each other. 
In fact, the BBN constraints require that M-sector must have smaller 
temperature than O-sector, $T' < T$. In this way, the contribution of 
mirror degrees of freedom to the Hubble expansion rate,  
equivalent to an effective number of extra neutrinos 
$\Delta N_\nu = 6.14 \cdot x^4$,  where $x = T'/T$,  
can be rendered small enough. E.g. the bound 
$\Delta N_\nu < 0.4$  implies $x < 0.5$, 
and for $x = 0.3$ we have $\Delta N_\nu \simeq 0.05$. 
This can be achieved by demanding that \cite{BCV}:

(A) at the end of inflation the O- and M-sectors  are (re)heated 
in an non-symmetric way, $T_R > T'_R$,  which can naturally occur  
in the context of certain inflationary models; 

(B) after (re)heating, at $T < T_R$, the possible 
particle processes between O- and M-sectors are too slow 
to establish equilibrium between them,  
so that both systems evolve adiabatically and the 
temperature asymmetry $T' /T$  remains nearly constant in all subsequent 
epochs until the present days. 

Mirror baryons, being invisible in terms of the ordinary photons, 
could constitute a viable dark matter candidate \cite{BCV}-\cite{BDM},   
and this possibility could shed a new light to the 
baryon and dark matter coincidence problem. 
First, 
the M-baryons have the same mass as the ordinary 
ones, $m'_{\rm B} = m_{\rm B}$. 
And second, the unified mechanism can be envisaged which generates 
the comparable baryon asymmetries in both O- and M-sectors,   
via $B-L$ violating scattering processes that transform 
the ordinary particles into the mirror ones. It is natural that 
these processes violate also CP due to complex coupling constants. 
And finally, their departure from equilibrium  is already implied 
by the above condition (B).  Therefore, all three Sakharov's 
conditions \cite{Sakh} for baryogenesis can be naturally satisfied. 
In addition, the mirror baryon density can be generated 
somewhat bigger than that of ordinary baryons,  $n'_{\rm B} \geq n_{\rm B}$, 
since the mirror sector is cooler than the ordinary one  
and hence the out-of-equilibrium conditions should be better fulfilled 
there.

\subsection{Neutrino as a Bridge to Mirror World } 

Let us take, for the simplicity,  the SM gauge symmetry 
$G = SU(3) \times SU(2) \times U(1)$ 
to describe the O-sector that contains the Higgs doublet $\phi$, 
and quarks and leptons: the left isodoublets   
$q_{Li} = (u_L, d_L)_i $, $l_{Li} = (\nu_L, e_L)_i$ 
and the right isosinglets   
$u_{Ri}$,  $d_{Ri}$, $e_{Ri}$ ($i=1,2,3$ is a family index).  
As usual, we assign a global lepton charge   
$L = 1$ to leptons and a baryon charge $B = 1/3$ to quarks.
If $L$ and $B$ were exactly conserved then 
the neutrinos would be massless and the proton would be stable.

But $L$ and $B$ are not perfect quantum numbers.
They are related to accidental global symmetries possessed by the 
SM Lagrangian at the level of  renormalizable couplings, 
which however can be explicitly broken by higher order operators 
cutoff by large mass scales $M$. 
In particular, 
D=5 operator $\cO \sim (1/M)(l\phi)^2$ ($\Delta L=2$), 
yields, after inserting the Higgs VEV $\langle \phi\rangle =v$,
small Majorana masses for neutrinos, $m_\nu \sim v^2/M$.

As for the M-sector, it should have 
a gauge symmetry $G' = SU(3)' \times  SU(2)'\times U(1)'$,
with a mirror Higgs doublet $\phi'$, and mirror quarks and leptons 
$q'_{Li} = (u'_L, d'_L)_i $, $l'_{Li} = (\nu'_L, e'_L)_i$;  
$u'_{Ri}$,  $d'_{Ri}$, $e'_{Ri}$,  
where the global charges $L'=1$ can be assigned to mirror leptons 
and $B'=1/3$ to mirror quarks. The mirror neutrinos get masses 
via $\Delta L'=2$ operator $\cO' \sim (1/M)(l'\phi')^2$.  
However, there can exist also mixed  gauge invariant operator 
$\cO^{\rm mix} \sim  (1/M) (l \phi)(l' \phi')$  
($\Delta L=1$, $\Delta L'=1$), 
that gives rise to the mixing between the
ordinary and mirror neutrinos \cite{FV}. 

In fact, all these operators can be induced by the same seesaw mechanism.  
Let us introduce $n$-species of the heavy Majorana neutrinos $N_a$, 
with the large mass terms $M g_{ab} N_a N_b$,
where $M$ is the overall mass scale and the matrix $g_{ab}$ of 
dimensionless Yukawa-like constants ($a,b=1,2,...,n$) 
taken diagonal without lose of generality. 
Then, as far as $N_a$ are the gauge singlets, they would couple 
the ordinary leptons $l_i=(\nu,e)_i$ and mirror leptons $l'=(\nu',e')_i$ 
with the equal rights: 
$y_{ia}l_i N_a \phi + y^\prime_{ia}\lpr_i N_a \phpr$. 
In this way, the heavy Majorana neutrinos 
play the role of messengers between ordinary
and mirror sectors.  Integrating them out in a seesaw manner, 
we obtain all relevant operators: 
\be{op-nu}
\cO = \frac{A_{ij}}{M} (l_i  \phi)( l_j\phi)  , ~~~ ~  
\cO' = \frac{A'_{ij}}{M} (l'_i \phi') (l'_j \phi') ,  ~~~~
\cO^{\rm mix} =  \frac{D_{ij}}{M} (l_i \phi) (l'_j \phi') ,     
\end{equation}
with the coupling constant matrices  
$A=y g^{-1}y^T$,  $A'=y'g^{-1}y^{\prime T}$  and 
$D=y g^{-1}y^{\prime T}$. 

In addition, by imposing mirror parity under the exchange $N_a \to N_a$, 
$l_{li} \to \tl'_{Ri} = C \ov{l}_{Li}$,  $\phi \to \phi^{\prime\ast}$,  
the Yukawa constant matrices in two sectors are related as 
$y' = y^\ast$, from which also stems that $A' = A^\ast$ and $D=D^\dagger$. 
Nevertheless, in the following in all formulas we keep constants 
$y$ and $y'$ for without specifying these relations. 

The interactions mediated by heavy neutrinos $N$ induce, along with the 
$\Delta L =2$ processes $l\phi \to \barl\barphi$, etc. in O-sector and their 
$\Delta L' =2$ analogues in  M-sector, also 
scattering processes $ l_i \phi \to \bar \lpr_k \barphpr $ etc. 
($ \Delta L=1,  \Delta L'=1$)
that transform O-particles into their 
M-partners. It is easy to see that they all three conditions 
for baryogenesis \cite{Sakh,KRS} are naturally  fulfilled: 

(i) $B-L$ violation is automatic: these processes violate 
$L(L')$  while conserve $B(B')$, and thus both $B-L$  and $B'-L'$ 
are violated. 

(ii) CP violation in these processes occurs due to 
due to complex Yukawa matrices $y$ and $y'$. 
As a result, cross-sections with leptons and anti-leptons in the initial state
are different from each other. It is important to stress, that CP-asymmmetry 
emerges in $\Delta L=1$ processes as well as in $\Delta L=2$ processes, 
due to the interference between the tree-level and one-loop diagrams 
shown in ref. \cite{BB-PRL}.  The direct calculation gives:
\begin{eqnarray}\label{CP}
&&
\sigma (l\phi\to \barlpr\barphpr) -
\sigma(\barl\barphi \to \lpr\phpr) =
(- \Delta\sigma  - \Delta\sigma' ) /2
\, ,  \nonumber \\
&&
\sigma (l\phi\to \lpr\phpr) -
\sigma(\barl\barphi \to \barlpr\barphpr) =
( -\Delta\sigma + \Delta\sigma' )/2
\, ,   \nonumber  \\
&&
\sigma (l\phi\to \barl\barphi) -   
\sigma(\barl\barphi \to l\phi) = \Delta\sigma \, , 
\end{eqnarray}
where 
\begin{eqnarray}\label{J}
&&
\Delta\sigma = {{3J\, S} \over {32\pi^2 M^4}} \, , ~~~~~
J= {\rm Im\, Tr} [ g^{-1}(y^{\dagger}y)^\ast g^{-1}
(y^{\prime\dagger}y^\prime) g^{-2} (y^\dagger y) ]  \, , \\
&&
\Delta\sigma' = {{3J'\, S} \over {32\pi^2 M^4}} \, , ~~~~~
J'= {\rm Im\, Tr} [g^{-1} (y^{\prime\dagger}y^\prime)^\ast g^{-1}
(y^{\dagger}y) g^{-2} (y^{\prime\dagger}y^\prime)]  
\end{eqnarray}
with $S$ being the c.m.\ energy square,  and $J$ and $J'$ the CP-violation 
parameters (notice that $J'$ is obtained from $J$ by exchanging  
$y$ with $y^\prime$). \footnote{
It is interesting to note that the tree-level amplitude for the 
dominant channel $l\phi\to \barlpr\barphpr$ goes as $1/M$ and the
radiative corrections as $1/M^3$.
For the channel $l\phi\to \lpr\phpr$ instead, both tree-level and
one-loop amplitudes go as $ 1/M^2$.
As a result,  the cross section CP asymmetries are comparable 
for both $l\phi\to \barlpr\barphpr$ and $l\phi\to \lpr\phpr$ channels. }

(iii) It is essential that  $\Delta L=1$ processes $ l\phi \to \bar \lpr\barphpr $,
as well as $\Delta L=2$ ones like $l\phi \to \barl\barphi$, $ll\to \phi\phi$, 
stay out of equilibrium.  (Notice, that for the first reaction this is required 
also in view of the BBN constraints, see condition (B).) 
Thus their rates,  correspondingly 
$\Ga_{1,2} = Q_{1,2}n_{\rm eq}/8\pi M^2$ ,  where  
$Q_1={\rm Tr}(D^\dagger D) = 
{\rm Tr}[g^{-1}(y^{\prime\dagger}y^\prime)^\ast g^{-1} (y^\dagger y)]$ 
and  $Q_2= 6 {\rm Tr}(A^\dagger A) =
6 {\rm Tr}[g^{-1}(y^{\dagger}y)^\ast g^{-1}(y^\dagger y) ] $   
(the sum is taken over all isospin and flavour indices of the 
initial and final states), 
%
and $n_{\rm eq}\simeq (1.2/\pi^2)T^3$ is an equilibrium density  
per one bosonic degree of freedom, should not exceed much 
the Hubble parameter $H = 1.66 \, g_\ast^{1/2} T^2 / M_{Pl} $
for temperatures $T \leq T_R$, where 
$g_\ast$ is the effective number of particle degrees of freedom, 
namely $g_\ast \simeq 100$ in the SM. 
In other words, the dimensionless parameter 
\be{K12} 
k = \left({{\Gamma_1 + \Gamma_2} \over {H}}\right)_{T=T_R} 
\simeq 3 \times 10^{-3}\, {{(Q_1+Q_2) T_R M_{Pl}} \over {g_\ast^{1/2}M^2}}  
\ee
should not be much larger than 1. Namely, the energy 
density transferred from ordinary to mirror sector will be crudely
$ \rho^\prime \approx (8 k_1/g_\ast)\rho$ \cite{BB-PRL}. 
Thus, translating this to the BBN limits, 
this corresponds to a contribution equivalent to an 
effective number of extra light neutrinos 
$\Delta N_\nu = 6.14 x^4 \approx k/14$. Therefore, the BBN 
bounds imply a weaker limit $k < 7$, while the stronger 
constraint  $k < 1.5$ or so comes 
from the large scale structure limit $x<0.3$ \cite{M-cosm}.



\subsection{Leptogenesis between O- and M-worlds:  
$\Om'_{\rm B}/\Om_{\rm B}$ from $n'_{\rm B}/ n_{\rm B}$} 

The leptogenesis scheme \cite{BB-PRL,Alice} 
which we discuss now is based on scattering
processes like $l\phi \to l'\phi'$ mediated by heavy Majorana neutrinos $N$ 
rather than on their decay $N\to l \phi$. 
A crucial role in our considerations is played by the reheating
temperature $T_R$, at which the inflaton decay and entropy production
of the Universe is over, and after which the Universe is dominated by
a relativistic plasma of ordinary particle species.
As we discussed above, we assume that after the postinflationary
reheating, different temperatures are established in the two
sectors: $T'< T$,  i.e. the mirror sector is cooler than
the visible one, or ultimately, even completely ``empty".
We also assume that the heavy neutrino masses $gM$ are larger than 
$T_R$ and thus cannot be thermally produced.
As a result, the usual leptogenesis mechanism 
via $N\to l\phi$ decays is ineffective.
Nevertheless, a net $B-L$ could emerge in the Universe as
a result of CP-violating effects in the unbalanced production
of mirror particles from ordinary particle collisions, $l\phi \to l'\phi'$. 

However, this mechanism would generate
the baryon asymmetry not only in the observable sector,
but also in the mirror sector. In fact, the two sectors are completely
similar, and have similar CP-violating properties.
We have scattering processes $l\phi \to l'\phi'$ which transform the 
ordinary particles into their mirror partners, and CP-violation effects in 
this scattering owing to the complex coupling constants.
These processes are most effective at temperatures 
$T\sim T_R$ but they are out of equilibrium.
In this case, at the relevant epoch, the ordinary observer should detect that  
(i) his world is losing entropy due to leakage of particles to the 
mirror sector 
(ii) leptons $l$ leak to the M-sector with different rate than anti-leptons $\bar{l}$
and so a non-zero $B-L$ is produced in the Universe. 
On the other hand, his mirror colleague would see that
(i) entropy production takes place in M-world, and 
 (ii) leptons $l'$ and anti-leptons $\bar{l}'$ appear with different rates.  
Therefore,  he also would observe that 
a non-zero $B'-L'$ is induced in his world.

One would naively expect that in this case the baryon asymmetries
in the O- and M-sectors should be literally equal,
given that the CP-violating factors are the same for both sectors.
However, we show that in reality, Baryon asymmetry  in the M sector,
since it is colder, can be about an order of magnitude bigger   
than in the O sector, as far as washing out effects are taken into account.
Indeed, this effects should be more efficient for the hotter O-sector
while they can be negligible for the colder M sector, and 
this could provide reasonable differences between the two worlds
in case the exchange process is not too far from equilibrium.
The possible marriage between dark matter and the leptobaryogenesis
mechanism is certainly an attractive feature of our scheme.    

The evolution of
the $B-L$ and $B'-L'$ number densities is determined by the CP asymmetries   
shown in eqs.~(\ref{CP}) and obey respectively the equations 
\be{L-eq-2}
{{ d n_{\rm BL} } \over {dt}} + (3H + \Gamma) n_{\rm BL} 
= \frac34 \Delta\sigma \, n_{\rm eq}^2 , ~~~ 
{{ d n'_{\rm BL} } \over {dt}} + (3H + \Gamma' ) n'_{\rm BL}  =
  {3 \over 4} \Delta\sigma' \, n_{\rm eq}^2   \; ,
\end{equation}
where $ \Gamma = kH$ and $\Gamma' = k x^3 H$ ($x=T'/T$) are 
respectively 
the effective total rates of the $\Delta L=1,2$ reactions for O-sector, and 
 $\Delta L'=1,2$ ones for M-sector.  
For the CP asymmetric cross section $\Delta\sigma$ we take
the thermal average c.m.\ energy square $S \simeq 17\, T^2$. 

Integrating this equations, we obtain for the final 
$B-L$ and $B'-L'$ asymmetries of the Universe, respectively  
$Y_{\rm BL} = n_{\rm BL}/s = D(k) \cdot Y_{0}$ 
and $Y'_{\rm BL} = n'_{\rm BL}/s = D(kx^3) Y_0$,  
where $s=(2\pi^2/45)g_\ast T^3$ is the entropy density, 
\be{BL}
Y_{0} \approx 2 \times 10^{-3} \, 
{{J\, M_{Pl} T_R^3} \over {g_\ast^{3/2} M^4 }}  \simeq  
10^{-10} \, \frac{Jk^2 }{Q^2} \left(\frac{T_R}{10^9~ {\rm GeV}}\right)   
\end{equation}
is a solution obtained in the perfect out-of-equilibrium limit  $k \to 0$ ($\Ga \ll H$), 
and the damping factor $D(k)$ has a form \cite{Alice}   
\be{Dk}
D(k) = \frac{3e^{-k}}{10k^4} \left[4k^3 -6k^2+6k-3 +3 e^{-2k}\right] 
 + \frac{6}{5k^3} \left[ 2 -(k^2 + 2k +2) e^{-k} \right] ,
\end{equation}
where the first and second  terms  
correspond to the integration of (\ref{L-eq-2}) respectively
in the epochs before and after reheating  ($T > T_R$ and $T < T_R$).

In the limit $ k \ll 1 $ one has  $ D(k) = 1 $. 
However, for $k\sim 1$ the depletion becomes reasonable 
(see. Fig. 5 in ref. \cite{Alice}):  
namely,  for $k=1,2$, respectively $D(k) = 0.35, 0.15$.
On the other hand, $k\gg 1$ the mirror sector will be 
heated too much which  would contradict to the 
BBN limit $k  <7$, while stronger limit $k < 1.5$ is 
from the large scale structure and CMB data \cite{M-cosm}. 
Therefore, even if $k\sim 1$, anyway $k' = x^3 k \ll 1$ due to the 
smallness of the temperature ratio $x = T'/T$ and so $D(kx^3) \approx 1$.

Now taking into the account that in both sectors the
$B-L$ densities are reprocessed into the baryon number densities
by the similar sphaleron processes \cite{KRS},
we have $Y_{\rm B} = aY_{\rm B-L} $ and $Y'_{\rm B} = aY'_{\rm B-L}$,
with coefficients $a$ equal for both sectors, 
and also baryon masses are the same, $m'_{\rm B} = m_{\rm B}$, 
we see that the cosmological densities of the ordinary 
and mirror baryons should be related as
\be{omegabp}
\frac{\Om'_{\rm B}}{\Om_{\rm B}} = 
\frac{m'_{\rm B}Y'_{\rm B}}{m_{\rm B} Y_{\rm B}} = 
\frac{D(kx^3)}{D(k)}
\end{equation}
If $k \ll 1$,
depletion factors in both sectors are $ D \approx D' \approx 1 $
and thus we have that the M- and O- baryons would have the
same densities, $\Omega'_{\rm B} \approx \Omega_{\rm B}$.
However, if $k \sim 1$, then we would have 
$ \Omega'_{\rm B} > \Omega_{\rm B} $,
and thus all dark matter of the Universe could be constituted 
of mirror baryons.
Namely, for $k \simeq1.5$ we would have 
that $\Omega'_{\rm B}/\Omega_{\rm B} \approx 5$.

\subsection{Breaking Mirror Parity: 
$\Om'_{\rm B}/ \Om_{\rm B}$ from  $m'_{\rm B} / m_{\rm B}$ } 

Let us conclude with  the following remark. 
In the estimation (\ref{omegabp}) we have assumed 
that $m'_{\rm B} = m_{\rm B}$. Then, for explaining the 
$\Om'_{\rm B} > \Om_{\rm B}$, we had to assume that $k\sim 1$. 
However, one could make twist of the approach and 
consider the possibility of having $k \ll 1$, but 
$m'_{\rm B} > m_{\rm B}$. This could occur 
if the mirror symmetry is spontaneously broken \cite{BDM}.

Indeed, Mirror parity $G\leftrightarrow G'$ tells that  
all coupling constants (gauge, Yukawa, Higgs) 
are identical for both sectors. If in  addition  
once the O- and M-Higgses have the equal VEVs, 
$\langle \phi\rangle = \langle \phi'\rangle =v$,  
then the mass spectrum of mirror particles, 
elementary as well as composite, 
would be exactly the same as that of ordinary ones, and 
so $m'_{\rm B} = m_{\rm B}$. 
  
However, there is no fundamental reason to think 
that the Nature does not apply 
the old principle "The only good parity ... is a broken parity" 
and mirror parity remains an exact symmetry.  
Namely, it could be spontaneously broken 
due to different VEVs of the Higgs doublets, 
$\langle \phi\rangle =v$ and $ \langle \phi'\rangle =v'$,
with $v'/v = \zeta$ different from 1 \cite{BDM}. 

Clearly, the weak gauge boson masses would scale as 
$M'_{W,Z} = \zeta M_{W,Z}$ and hence the weak interaction 
constant $G'_F$ in the M-sector would be $\zeta^2$ times smaller than 
the Fermi constant $G_F$. In addition, masses all elementary mirror 
fermions would scale by factor $\zeta$: e.g. $m'_e = \zeta m_e$ 
for mirror electron and $m'_{u,d} = \zeta m_{u,d}$ for the light mirror quarks. 

However, as far as strong interactions are concerned,
a big difference between the electroweak scales $v'$ and $v$   
will not cause the similar big
difference between the confinement scales in two worlds \cite{BDM}. 
Indeed, mirror parity is valid at higher (GUT) scales,
the strong coupling constants in both sectors would evolve down in energy
with same values until the energy
reaches the value of the mirror-top ($t'$) mass.
Below it $\alpha'_{s}$ will have a different slope than $\alpha_{s}$.  
It is then very easy to calculate the value of the scale $\Lambda'$
at which $\alpha'_{s}$ becomes large. 
The value $\La'/\La$ scales rather slowly with $\zeta = v'/v$, 
approximately as $\zeta^{0.3}$ \cite{BGG}. 
Therefore, 
taking $\Lambda=200$ MeV for the ordinary QCD, then for 
e.g. $\zeta\sim 100$ we find $\Lambda'\simeq 800$ MeV or so.
On the other hand, we have $m'_{u,d}=\zeta m_{u,d}$ so that
masses of the mirror light quarks $u'$ and $d'$
also get close to $\Lambda'$ but do not exceed it  
(situation similar between the $s$ quark mass $m_s$ and $\La$ 
in the ordinary QCD). 

So, one expects that for $\zeta \leq 100$ the mirror baryon 
mass scales as $m'_{\rm B}/m_{\rm B}\sim \La'\La \sim \zeta^{0.3}$, 
then one would get $m'_{\rm B} \sim 5$ GeV or so, 
while the electron which scales as $m'_e/m_e \simeq \zeta$ 
one would have $m'_e \simeq 50$ MeV.  This situation looks 
interesting, since apart of the possibility to providing 
$\Om'_{\rm B} /\Om_{\rm B} \simeq 5$ even for $k\ll 1$, 
it would also imply that mirror atoms are much more compact 
than the ordinary atoms: 
their Bohr radius scales as $r'_H \simeq \zeta^{-1} r_H$, 
and thus mirror matter should be much less collisional and dissipative 
than the ordinary one. Also, the hydrogen  recombination  
and photon decoupling in M-sector would occur much 
earlier the matter-radiation equality epoch, and as 
a consequence, mirror matter will manifest rather like a 
cold dark matter. 

As for the mirror nucleons: protons and neutrons 
their masses both scale roughly as $\zeta^{0.3}$ 
with respect to the usual nucleons.
However,  since the light quark mass difference scales 
$(m'_d-m'_u)\approx \zeta (m_d-m_u)$
we expect the mirror neutron $n'$ to be heavier than the mirror
proton $p'$ by few hundred MeV. 
Clearly, such a large mass difference cannot be compensated by
the nuclear binding energy and hence
even bound neutrons will be unstable
against $\beta$ decay $n'\to p' e' \bar{\nu}'_e$.
Thus in such an asymmetric mirror world hydrogen will be the only stable 
nucleus \cite{BDM}.


\begin{theacknowledgments}
I would like to thank Carlos Munoz and other organizers of the 
"Dark Side of the Universe" (DSU 2006) for the hospitality in Madrid. 
 The work is partially supported by the MIUR biennal grant
for the Projects of the National Interest PRIN 2004 on "Astroparticle Physics".  
\end{theacknowledgments}


\end{document}